# DLOT: An open-source application to assist human observers

Ashwin T S[a*], Danish Shafi SHAIKH[a], & Ramkumar RAJENDRAN[a]
[a]*Indian Institute of Technology, Bombay*
*ashwindixit9@gmail.com

**Abstract:** Adaptive intelligent educational systems are gaining popularity, offering personalized learning experiences to students based on their individual needs and styles. One crucial feature of such systems is real-time personalized feedback. However, identifying real-time learning processes impacting student performance remains challenging due to data volume constraints. Current research often relies on labor-intensive human observation, which is time-consuming and not scalable. To efficiently collect real-time data, an observation tool is essential. Qualitative/Mixed Method research explores participant experiences in education, social science, and healthcare, utilizing methods like focus groups and observations. However, these methods can be labor-intensive, particularly in maintaining observation time intervals. Existing tools lack comprehensive support for education-focused focus groups and observations. To address these issues, this paper introduces the Data Logging and Organizational Tool (DLOT), a flexible tool designed for qualitative studies with human observers. DLOT offers customizable time intervals, cross-platform compatibility, and data saving and sharing options. The tool empowers observers to log timestamped data and is available on GitHub. The DLOT was validated through two studies. The first study predicted students' affective states using real-time annotations collected via DLOT, observing 30 students in each class. The second study created multimodal datasets in a computer-enabled learning environment, observing 38 students individually. A successful usability test was conducted, offering a potential solution to challenges in real-time learning process identification and labor-intensive qualitative research observation.

## 1. Introduction

Adaptive intelligent educational systems are becoming increasingly popular, particularly in distance learning and online education. These systems are designed to provide personalized learning experiences to students, catering to their individual learning needs and styles. One of the key features of such systems is the provision of personalized feedback to students based on their learning processes. To be effective, this feedback needs to be provided in real-time as the student interacts with the learning environment (Mousavinasab et al., 2021, Ashwin & Guddeti, 2020, Taub et al. 2021).
Identifying the various learning processes that contribute to a student's real-time performance is a challenging task. These processes include self-regulated learning, cognitive engagement, metacognition, affective states, behavioral and interactional dynamics with peers, collaborative learning, uncertainty, feedback seeking, empathy, and group dynamics. To train machine learning models to detect these processes in real-time, a large amount of annotated data is necessary to identify, classify, or predict these constructs (Sinatra, Heddy, & Lombardi, 2015, Praharaj *et al.* 2021, Ashwin & Guddeti, 2019, Munshi *et al.* 2018).

Current research in this area involves human observation, where video recordings are analyzed after the study has concluded. However, this method is time-consuming, labor-intensive, and not scalable. Automated machine learning models also face challenges in real-time identification of these constructs due to insufficient data (Wu *et al.* 2022, TS & Guddeti, 2020). Hence, there is a need for a reliable observation tool that can collect data quickly and efficiently in real-time. Such a tool could revolutionize the field of adaptive intelligent educational systems, providing personalized feedback to students based on their real-time learning processes, leading to improved learning outcomes.

On the other hand, Qualitative/Mixed Method research is an approach to research that focuses on exploring the experiences, attitudes, and perspectives of participants in a study. It is often used in education, social science, and healthcare research, where understanding human behavior is a key objective. The research involves collecting data through a variety of methods, such as questionnaires, interviews, focus groups, tests, observations, and secondary data (Johnson, Burke, & Turner, 2003). While some methods like questionnaires and tests, can be conducted without human observation, others, such as focus groups and observations, require human observers to gather data. These methods are crucial in providing insights into human behavior, emotions, and experiences. However, the process of human observation can be labor-intensive, especially when there are a large number of subjects or the observation period is lengthy. In addition, maintaining the exact time interval between observations can be challenging for human observers. An application or tool that automatically sets the timer and provides a prompt for the observer to log the data would be incredibly useful in such cases.

To facilitate the collection of data with specific labels, we have developed this app, drawing inspiration from the Human Affect Recording Tool (HART). Similar to various tools commonly employed in qualitative and mixed-method research, such as Teamscope, Open Data Kit (ODK), and KoboToolbox, which find utility in questionnaires, tests, and secondary data collection, our app serves a comparable purpose. It's worth noting that while these established tools are often used for activities like questionnaires, tests, and secondary data collection, they may not be ideally suited for tasks like focus groups and observations. Additionally, it's important to emphasize that not all these tools are available as free and open-source options. Furthermore, the adaptation of new technology might involve a learning curve for both participants and researchers, potentially influencing the efficiency and accuracy of the data collection process. In the context of educational research, the Human Affect Recording Tool (HART) has been utilized for Quantitative Field Observations (QFOs), primarily focusing on monitoring students' affect and behavior (Ocumpaugh et al. 2015). However, in response to this limitation, our innovation aims to empower human observers by providing a mechanism to log data along with timestamps.

Hence, this paper contributes to a DLOT (Data Logging and Organizational Tool) developed to aid qualitative and also quantitative studies involving human observers. DLOT is flexible and can be customized to set time intervals, works with different operating systems, works for the individual, team (or group), different sets of codes/labels, and offers various options for saving and sharing data.

Moreover, it's pertinent to emphasize that the central focus of this article is to present a concise and comprehensive overview of the Data Logging and Organizational Tool (DLOT). Rather than a typical research study, the article is tailored to provide an in-depth description of DLOT's features and functionality. The rest of the paper is organized as follows: Section 2 provides details of DLOT, Section 3 discusses the validation, and Section 4 concludes the paper with future directions.

## 2. DLOT

The DLOT tool was developed using Figma and React Native. Data export was handled using the docx and xlsx packages. State management was implemented using useState hooks, and dynamic runtime changes were handled with the useEffect hook. The app development and testing were done using Expo. Additional details about the DLOT tool development are mentioned in Appendix A[1]. The complete details, including the readme information, can be found in the GitHub repository.

The DLOT tool offers a wide range of flexible options that cater to various research requirements:

*Class Labels or Categories*: Researchers using DLOT can create custom labels or categories tailored to their study objectives. For example, Figure 1a illustrates class labels

---
[1] https://github.com/danishsshaikh/DLOT/Appendix A

such as "engagement" to "on-task," which can be modified to suit different studies. For instance, in a metacognition study, labels can be defined as "planning," "monitoring," "evaluating," and more, based on the specific research context. Prior to using the DLOT tool, researchers need to establish a set of categories or class labels that will be used throughout the study. This step involves defining the specific labels that will be assigned to observations. Once set, these categories remain consistent and cannot be changed during the course of the study.

*Time Flexibility*: DLOT offers researchers the flexibility to set the observation timer according to their study requirements. By default, the timer is set to 10 seconds, providing a suitable time frame for observations. The timer duration is conveniently displayed atop Figure 1a.

*Selection of Categories*: DLOT empowers researchers to design observation interfaces that include both radio buttons and checklists, as demonstrated in Figure 1a. This feature enables researchers to make multiple selections based on their study's specific needs, enhancing the richness and versatility of data collection.

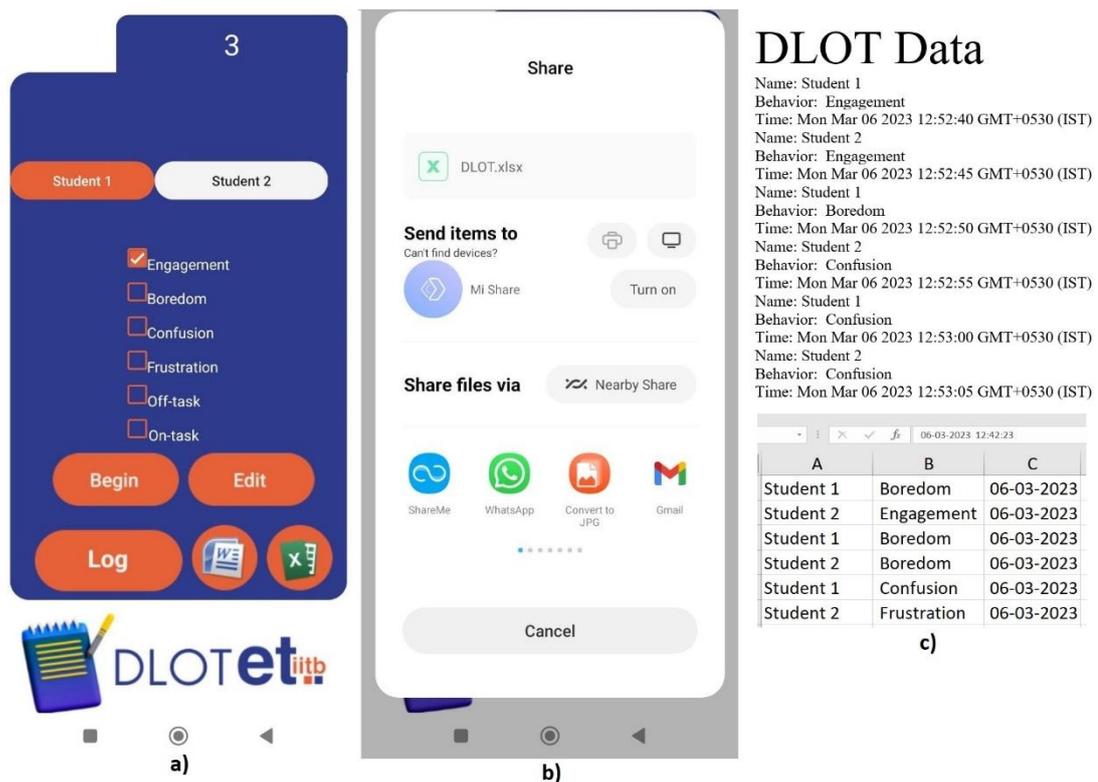

*Figure 1.* There are three screenshots attached that illustrate the data logging and organizational tool. The first screenshot (Figure a) displays the interface where the student's name, class labels, timer, and a log button can be found. Once the logging process is completed, the data can be downloaded in either xlsx or docx format. After selecting the desired format, the data can be stored locally on the mobile device or in any other preferred location, as depicted in Figure b). In Figure c), a sample of the logged data is displayed in a particular format. In this format, the student's name or ID acts as the primary key, and the data is organized based on the timestamp.

*Adjusting Observations (Number of Students and Group Size)*: DLOT allows researchers to customize the number of students observed, accommodating studies involving individuals, small teams, or larger groups like classrooms. Researchers can adapt the tool to their specific research setting.

*Platform Compatibility and Data Storage*: DLOT is compatible with both Android and iOS devices, ensuring broad accessibility for researchers using different operating systems. Additionally, the tool supports multiple data storage formats, including text and xlsx. This versatility facilitates easier data analysis and sharing, streamlining the research workflow.

*File Sharing Options*: Once data logging is complete, users can choose to store the final data locally or share it through various options such as Google Drive, WhatsApp, Telegram, Bluetooth, LINE, etc. Figure 1b illustrates the convenient file sharing functionality of DLOT.

*Open-Source Availability*: DLOT is available as an open-source application with its GitHub code[2] and official webpage guidelines and read-me information[3]. Sample screenshots of DLOT are shown in Figure 1. The tool also includes a discussion forum on GitHub where researchers can ask questions about using the app or modifying default values, among other topics.

*Training Observers for Data Annotation*: To utilize the DLOT tool effectively, human observers need to undergo training in annotating or classifying data based on standard definitions or study guidelines. For instance, in a basic emotion detection study, multiple observers should observe the same student(s) and provide labels following established guidelines, such as a facial action coding system. Observers can choose various annotation methods, including annotating together or independently and later discussing any disagreements. Importantly, observers must adhere to the specific purpose of the observation. For example, suppose the annotations aim to train a classification model like convolutional neural networks. In that case, guidelines should emphasize labeling based on the specific instance observed rather than relying on prior frames or external sources of knowledge.

The following steps should be followed to use the DLOT:
1. Train the human observers on the required class labels or categories and establish inter-rater reliability.
2. Install the application.
3. Check if the required parameters are present in the application.
4. If yes, go to step 5; Else, modify the open-source code according to the requirements.
5. Set the parameters and collect data.
6. Save the data. (More details are in Appendix B[4])

By embracing flexibility, customization, and compatibility, DLOT empowers researchers to efficiently collect and analyze qualitative data in a manner that suits their specific study objectives and research settings.

## 3. Validation

The Data Logging and Organizational Tool (DLOT) was validated using two studies. The first study focused on affective state prediction in a classroom setting, where real-time annotations were collected using DLOT. A total of 30 students were observed in each class, and their affective states, including engaged, boredom, confusion, frustration, and neutral, were predicted. The timer interval was set to 5 seconds as per the study requirement. The second study incorporated DLOT as a component in generating multimodal datasets (interaction logs, screen recordings, human observations and facial expressions), which involved 38 participants in a computer-enabled learning environment. This study involved observing a single student at a time, with a time interval of 10 seconds throughout the study. Three observers used the tool in the first study, while five used it in the second. The validation process encompassed an evaluation of system usability, conducted via the System Usability Scale (SUS) analysis, yielding an average score of 93. SUS, a widely utilized questionnaire, employs a 5-point Likert scale and subsequently transforms the scores to a range of 0 to 100 (Brooke, 1995). The attained score of 93 reflects the collective assessment by participants and underscores the commendable usability of the system. Observers reported that the tool was handy and saved them significant time. The user interface was deemed easy to navigate, and the installation process was straightforward.

DLOT was tested on different operating systems, including Android and iOS, and performed well without technical glitches or issues. The open-source code of DLOT was made available for observers. One observer from the study one used it and found it easy to modify and make changes. He also mentioned the information provided in the open-source code was clear and concise, and the accompanying discussion forum proved helpful in addressing queries and concerns.

---

[2] https://github.com/danishsshaikh/DLOT
[3] https://danishsshaikh.github.io/DLOT/
[4] https://github.com/danishsshaikh/DLOT/Appendix B

During the studies, DLOT was used continuously for extended periods while observing 30 students, and no technical issues such as hanging or data loss were experienced. Overall, the validation process and user feedback demonstrate the reliability, usability, and effectiveness of the DLOT tool in diverse educational contexts.

Since there are not many manual note-taking tools in the education domain, it is relevant to compare DLOT with HART, the only existing tool that is similar to DLOT in terms of observing students and noting down their states with a time stamp. The comparison and advantages of DLOT over HART are as follows: *Fixed Class Labels*: HART has predefined class labels or categories based on Baker Rodrigo Ocumpaugh monitoring protocol (BROMP) (Ocumpaugh, 2015), limiting the customization options for researchers. *Time Limitations*: HART imposes a fixed countdown clock of 3 minutes for each observation, which restricts the coding time to 20 seconds for BROMP observers. *Single Selection*: HART only allows the selection of one class label or category per student within a time interval, using radio buttons. *Platform Compatibility*: HART is designed exclusively for the Android platform, limiting its accessibility to researchers. *Data Storage Formats*: HART stores data only in the txt format, which may limit the options for data analysis and sharing. *File Sharing Options*: HART data files are stored locally on Android devices and can be shared via email or transferred to computers using USB cables. As explained in the previous section, DLOT offers advantages compared to HART in terms of flexibility in labeling, time management, selection options, platform compatibility, data storage formats, and file sharing options.

Limitations: In this study, there are several limitations of the Data Logging and Organizational Tool (DLOT). While DLOT demonstrates potential for real-time qualitative studies, a comparative usability analysis with existing tools was not conducted. Additionally, the presence of observers using DLOT could introduce bias, demanding caution. Security and privacy concerns inherent to digital platforms must also be considered. Furthermore, DLOT's reliance on smartphone technology and internet access might restrict its use in certain contexts.

In addition to its use in the education sector, this tool has the potential to be utilized in multiple other domains. For instance, in research studies, researchers can use the app to log data regarding participant behaviors, reactions to stimuli, or experiment results. Clinical psychologists can also use the app to record data concerning patient sessions, including symptoms, behaviors, or progress over time. Moreover, coaches or athletes can utilize the tool to log data related to training sessions or competitions, such as times, distances, or personal bests.

## 4. Conclusion

The Data Logging and Organizational Tool (DLOT) is a flexible and user-friendly application developed for data collection and analysis in educational research. It offers researchers the ability to create custom class labels or categories tailored to their study objectives, adjust observation timers, and choose from multiple selection options for data collection. DLOT is compatible with Android and iOS platforms, ensuring accessibility for researchers using different operating systems. It supports various data storage formats, facilitating easy analysis and sharing of data. The tool also provides convenient file sharing options, allowing users to store data locally or share it through different platforms. Being an open-source application, DLOT empowers researchers to modify and customize the tool according to their specific needs. The GitHub repository provides comprehensive information and a discussion forum for user support. Validation studies conducted on affective state prediction and multimodal dataset creation have shown the reliability and usability of DLOT. Observers reported that the tool was highly useful and time-saving, with an intuitive interface and straightforward installation process. DLOT performed well on different operating systems, ensuring its effectiveness across platforms.

Several future directions for the DLOT include adding data visualization capabilities, user authentication, cloud storage integration, customization in different sets of options, integration with other educational apps, and audio and video capabilities for recording and analyzing speech for unstructured data.